# Anomalous Interlayer Exciton Diffusion in Twist-Angle-Dependent Moiré Potentials of $WS_2$-$WSe_2$ Heterobilayers

Long Yuan[1], Biyuan Zheng[2], Jens Kunstmann[3], Thomas Brumme[4], Agnieszka Beata Kuc[5,6], Chao Ma[2], Shibin Deng[1], Daria Blach[1], Anlian Pan[2], and Libai Huang[1*]

[1]Department of Chemistry, Purdue University, West Lafayette, IN 47907, USA

[2]Key Laboratory for Micro-Nano Physics and Technology of Hunan Province, College of Materials Science and Engineering, Hunan University, Changsha 410082, People's Republic of China

[3]Theoretical Chemistry, Department of Chemistry and Food Chemistry, TU Dresden, 01062 Dresden, Germany

[4]Wilhelm-Ostwald-Institute for Physical and Theoretical Chemistry, Leipzig University, 04103 Leipzig, Germany

[5]Helmholtz-Zentrum Dresden-Rossendorf, Abteilung Ressourcenökologie, Forschungsstelle Leipzig, Permoserstr. 15, 04318, Leipzig, Germany

[6]Department of Physics & Earth Science, Jacobs University Bremen, 28759 Bremen, Germany

**Abstract:** The nanoscale periodic potentials introduced by moiré patterns in semiconducting van der Waals (vdW) heterostructures provide a new platform for designing exciton superlattices. To realize these applications, a thorough understanding of the localization and delocalization of interlayer excitons in the moiré potentials is necessary. Here, we investigated interlayer exciton dynamics and transport modulated by the moiré potentials in $WS_2$-$WSe_2$ heterobilayers in time, space, and momentum domains using transient absorption microscopy combined with first-principles calculations. Experimental results verified the theoretical prediction of energetically favorable K-Q interlayer excitons and unraveled exciton-population dynamics that was controlled by the twist-angle-dependent energy difference between the K-Q and K-K excitons. Spatially- and temporally-resolved exciton-population imaging directly visualizes exciton localization by twist-angle-dependent moiré potentials of ~100 meV. Exciton transport deviates significantly from normal diffusion due to the interplay between the moiré potentials and strong many-body interactions, leading to exciton-density- and twist-angle-dependent diffusion length. These results have important implications for designing vdW heterostructures for exciton and spin transport as well as for quantum communication applications.

Van der Waals (vdW) heterostructures assembled from graphene, hexagonal boron nitride, and transition metal dichalcogenide (TMDCs) have emerged as a new class of materials in exploring new quantum phenomena with designed functionalities[1-5]. Spatially-indirect interlayer excitons can be formed with electrons and holes localized in different TMDCs layers[6-9], with much longer lifetimes than the direct intralayer excitons, achieving long-range exciton- and valley-spin transport[3,4]. Further, lattice mismatch or rotational misalignment in the semiconducting TMDC heterostructures leads to the formation of moiré potentials (e.g. nanoscale periodic energy potentials for excitons), providing a configurable solid-state excitonic analog to ultracold atoms in optical lattices or photons in photonic crystals[10,11]. Scanning tunneling microscopy measurements and density functional theory (DFT) simulations for $MoS_2$-$WSe_2$[10,12,13] have shown that the bandgap of the heterobilayer vary spatially within the moiré pattern, suggesting amplitudes of moiré potentials as large as hundreds of meV. More recently, far-field optical measurements have indicated the existence of moiré excitons in semiconducting TMDC heterostructures by observations, such as trap exciton emission, flat exciton band, spatially varying valley polarization, and resonantly hybridized excitons[14-19].

How the quantum states of many-body systems are localized or delocalized at well-defined positions in space by energy potentials is a long-standing question in physics[20]. Although the photoluminescence (PL) and absorption spectroscopy measurements from aforementioned studies[14-19] provided indirect evidences for the localized excitons in moiré potentials, it still remains largely unknown the time- and length-scale of the localization. The transport of spatially-indirect excitons has been extensively investigated in electrically-biased coupled quantum wells, demonstrating long-range transport and quantum many-body effects, including Bose–Einstein condensation and superfluidity of excitons[21-23]; although moiré patterns do not exist in these systems. Recently, long-range interlayer exciton diffusion over micrometer lengthscale in TMDC heterostructures at both low temperature and room temperature was reported[3,24]; however, the role of moiré potentials has not been addressed. Here, we report direct ultrafast spatial imaging of interlayer exciton transport in the $WS_2$-$WSe_2$ heterobilayers with two different twist angles (0° and 60°) with a temporal resolution of ~200 fs and a spatial precision of ~50 nm. In combination with first-principles calculations, these results provide a comprehensive picture of the localization and delocalization of the interlayer excitons in the presence of twist-angle-dependent moiré potentials.

We performed measurements on the $WS_2$-$WSe_2$ heterobilayers, prepared using a modified two-step chemical vapor deposition (CVD) method (see Methods and Fig. S1 for more details) that provides contamination-free and atomically-sharp interfaces uniform over the micrometer length scale[25]. Fig. 1a shows an optical image of a typical $WS_2$-$WSe_2$ heterobilayer, grown on a Si wafer with a 285-nm-thickness oxide. The larger bottom layer is a single layer $WS_2$ (1L-$WS_2$), while the smaller top layer is a single layer $WSe_2$ (1L-$WSe_2$). The presence of the vertical heterostructures constituted of $WS_2$ and $WSe_2$ monolayers was confirmed using Raman spectroscopy (Fig. S2). The heterobilayers have two stacking orientations with twist angles of $\theta = 0°$ and 60°, which are energetically favorable in the modified two-step CVD growth[25]. The twist angle can be readily determined by the relative orientation of the top and bottom triangles, because the orientation of each triangle is directly correlated with its microscopic crystal orientation[26,27]. We further confirmed the stacking orientation using the second harmonic generation (SHG) spectroscopy (Fig. S3). The SHG intensity of $WS_2$-$WSe_2$ (60°) is strongly suppressed due to the destructive interference of the second harmonic fields, while it is enhanced in $WS_2$-$WSe_2$ (0°) by constructive interference[28]. The existence of moiré patterns can be directly observed by high-resolution annular dark-field (ADF) scanning transmission electron microscopy (STEM) image (Fig. 1b for 60° and Fig.S4 for 0°). The lattice mismatch of the monolayers of the WS$_2$-WSe$_2$ heterobilayers is about 4%, as confirmed by the selected area electron diffraction images shown in Fig. S5. This mismatch leads to the moiré periodicity of ~8.5 nm (Fig. 1b and Fig. S4).

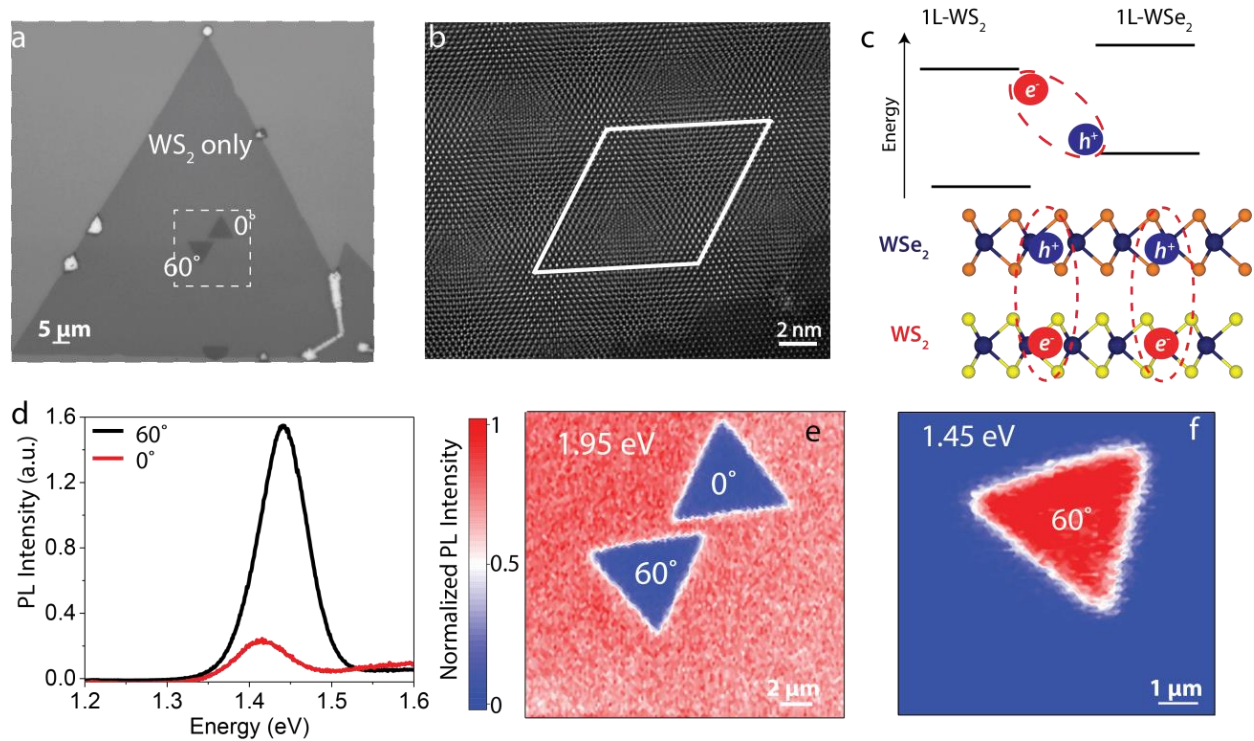

**Fig. 1 | Formation of the moiré superlattice and interlayer exciton emission. a**, Optical image of the CVD grown WS$_2$-WSe$_2$ heterobilayers with twist angles of 0° and 60°. **b**, High-resolution ADF STEM image of the WS$_2$-WSe$_2$ (60°). The moiré superlattice is marked by white solid lines showing a periodicity of ~8.5 nm, consistent with the mismatch of the lattice constants of the monolayers of ~4%. **c,** Schematic of the type-II band alignment in the WS$_2$-WSe$_2$ heterobilayers displaying interlayer exciton formation. **d**, PL spectrum of the WS$_2$-WSe$_2$ heterobilayers with twist angles of 0° and 60° at 78 K. The peak at ~1.45 eV corresponds to the interlayer exciton emission. PL spectra are normalized with respect to the peak intensity of the 1L-WS$_2$. **e**, PL image of the WS$_2$-WSe$_2$ heterobilayers with twist angles of 0° and 60° taken at 295 K. The detection energy is 1.95 eV, corresponding to the A exciton emission of the 1L-WS$_2$. **f**, PL image of the WS$_2$-WSe$_2$ heterobilayer with twist angle of 60° taken at 78 K. The detection energy is 1.45 eV, corresponding to the interlayer exciton emission.

The WS$_2$-WSe$_2$ heterobilayers have a type-II band alignment, as shown schematically in Fig. 1c[29,30], resulting in the formation of spatially-indirect interlayer charge-transfer excitons, with electrons and holes residing in the WS$_2$ and WSe$_2$ layers, respectively. As shown in Fig. 1d, the photoluminescence (PL) spectra at 78 K display a new emission peak at ~1.45 eV in both WS$_2$-WSe$_2$ with twist angles of 0° and 60°, which is attributed to the emission from the interlayer excitons[16]. Figs. 1e and 1f depict PL images collected at 1.95 eV and 1.45 eV, corresponding to the emission from the intralayer A exciton in WS$_2$ and the interlayer exciton emission, respectively. Significant PL quench of the WS$_2$ A exciton in the heterostructures indicates an efficient charge separation (Fig.1e). The PL microscopy image (Fig. 1f) of the interlayer exciton at 1.45 eV shows that the emission is homogenous over the heterostructure region, confirming the high quality of the interfacial contact. We also observed that the interlayer exciton emission intensity in the 60° heterobilayer was about one order magnitude higher than that of 0° (Fig. 1d).

To gain insights on how the energy landscape for the interlayer excitons is modified by the moiré pattern of the 0° and 60° heterobilayers, we performed first-principles calculations using DFT (see Methods and Supplementary Note 1 for details). Fig. 2a and Fig. 2b show a moiré supercell for the twist angles of 0° and 60°, respectively. Highlighted in color are different local, high-symmetry stacking configurations (Fig. S6); they are labeled according to the nomenclature introduced in Ref.[13], where $H_h^h$ and $R_h^M$ correspond to the atomic registries of 2$H$ and 3$R$ bulk polytypes, respectively. We calculated the energy of the four lowest-energy optical transitions between the valence-band maximum (VBM) in the K valley and the conduction-band minimum (CBM) in the K and the Q valleys (Fig. 2c and Table S1) for different stacking configurations. Notably, the lowest-energy transition is always K-Q and therefore K-Q interlayer excitons are expected to represent the ground state instead of the more commonly discussed K-K excitons. The moiré potentials plotted along the main diagonal of the moiré supercells, shown in Fig. 2d (also see Table S2), indicate that the spatial variations for 0° are much stronger (deep potential) than for 60° (shallow potential). The effect is illustrated for K-K singlet excitons by the vertical arrows in Fig. 2d. The gray area in Fig. 2d indicates the spatial variation of the energy difference between the K-Q and the K-K transitions, $\Delta E_{KQ}$ (also indicated in Fig. 2c). Averaged over the whole moiré pattern, the mean difference $\Delta E_{KQ}$ is larger for 0° than for 60° (88 meV vs. 62 meV, see Table S3).

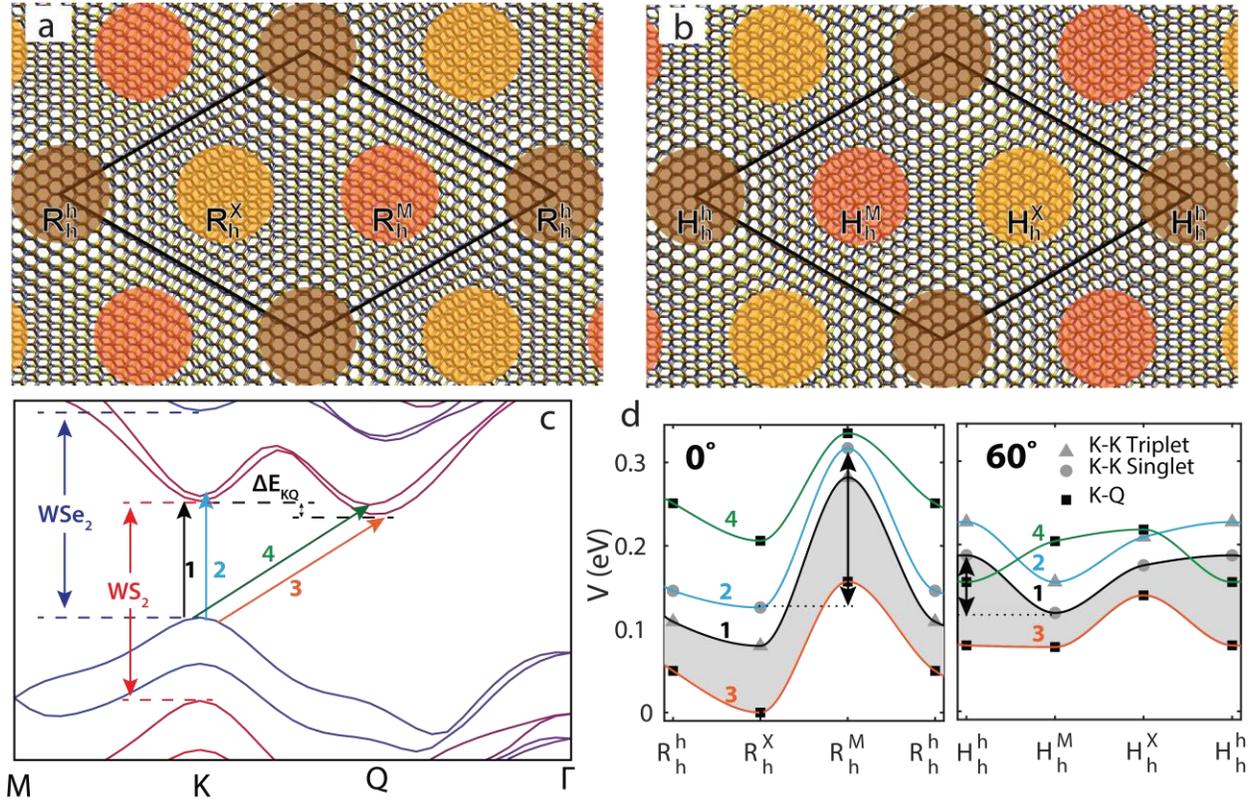

**Fig. 2 | Moiré potentials in WS$_2$-WSe$_2$ heterobilayers calculated with density functional theory. a**, **b**, Schematic of the moiré pattern of the WS$_2$-WSe$_2$ heterobilayers and their supercells (black) for twist angles of 0° and 60°, respectively. Brown, yellow, and orange circles mark regions with high-symmetry stacking configurations. **c**, Schematic representation of a typical band structure of a WS$_2$-WSe$_2$ heterobilayer in a (strained) primitive unit cell. The four lowest-energy transitions are indicated by arrows. They are either transitions in the K valley (K-K) or transitions between the valence band maximum in K and the conduction band minimum in the Q valley (K-Q). **d**, Approximate moiré potentials for the twist angles of 0° and 60° plotted along the main diagonal of the moiré supercells as marked in parts **a** and **b**, showing that the spatial variations in 0° are much stronger (deep potential) than in 60° (shallow potential) heterobilayers. The different lines correspond to the four lowest-energy optical transitions as marked in **c**. Circles and triangles for K-K indicate spin-singlet and spin-triplet excitations, respectively. Vertical arrows indicate the potential variations of the K-K spin-singlet excitons. The gray area indicates the difference between the lowest-energy K-Q and K-K transitions, $\Delta E_{KQ}$. Averaged over the whole moiré pattern, the difference $\Delta E_{KQ}$ is larger for 0° than for 60°. The theoretical results show that the moiré pattern leads to a twist-angle-dependent modulation of the energy landscape of the interlayer excitons.

Two predictions can be made based on the DFT calculations: (i) the population dynamics of the K-K and K-Q excitons should be affected by the twist-angle-dependent $\Delta E_{KQ}$; and (ii) the twist-angle-dependent moiré potentials should lead to different degrees of localization of the interlayer excitons in 0° and 60° heterobilayers.

First, to test the prediction (i), we selectively monitored electrons and holes in the K valleys using pump-probe transient absorption (TA) spectroscopy, as schematically shown in Fig. 3a. We employed a linearly-polarized pump excitation (at either 1.80 eV or 1.60 eV) to create A excitons in $WSe_2$. Following the photoexcitation, electrons are transferred from $WSe_2$ to $WS_2$ within 100 fs, while the holes remained in the $WSe_2$[31]. We used a linearly-polarized probe beam at 1.60 eV, to monitor the hole population in the K (or K') valley of $WSe_2$, and a probe energy of 2.00 eV, to track the electron population in the K (or K') valley of $WS_2$. Because the holes reside in the K valley of $WSe_2$ for both the K-K and K-Q interlayer excitons, the hole population probed at 1.60 eV corresponds to the sum of the K-K and K-Q exciton populations ($N_{K-K} + N_{K-Q}$). On the other hand, electrons are located in different valleys for the K-K and K-Q interlayer excitons. The dynamics probed at 2.00 eV reflects only the K-K interlayer exciton population ($N_{K-K}$), because only the electrons in the K valley of $WS_2$ contribute the TA signal. Thus, the population dynamics of the K-K and K-Q excitons can be elucidated by comparing the dynamics probed at 1.60 eV to that probed at 2.00 eV.

As shown in Fig. 3b, the decay time of ($N_{K-K} + N_{K-Q}$) in the heterostructures is ~1 ns at 295 K, which is much longer than the intralayer excitons in 1L-$WSe_2$ (~40 ps), due to the spatially-indirect nature of the interlayer excitons[6]. This lifetime increases to more than 3 ns at 78 K (Fig. 3c), showing similar temperature dependence as the PL decay (Fig. S7), likely due to the suppressed non-radiative recombination[8]. The hole dynamics do not exhibit significant twist-angle dependence (Fig. 3b and Fig. S8). The $N_{K-K}$ dynamics probed at 2.00 eV at 295 K is presented in Fig. 3d and Fig. S9, showing that the decay time for 0° is about 3 times shorter than that for 60°. Temperature-dependent measurements demonstrate that the K-K exciton decay time in both the 0° (Fig. 3e) and the 60° (Fig. S10) heterobilayers become significantly shorter at lower temperatures, in stark contrast to the total exciton population shown in Fig. 3c. All measurements were carried out at an exciton density of $4.1 \times 10^{12}$ cm$^{-2}$ and the dynamics do not show significant dependence on exciton density (Fig. S11).

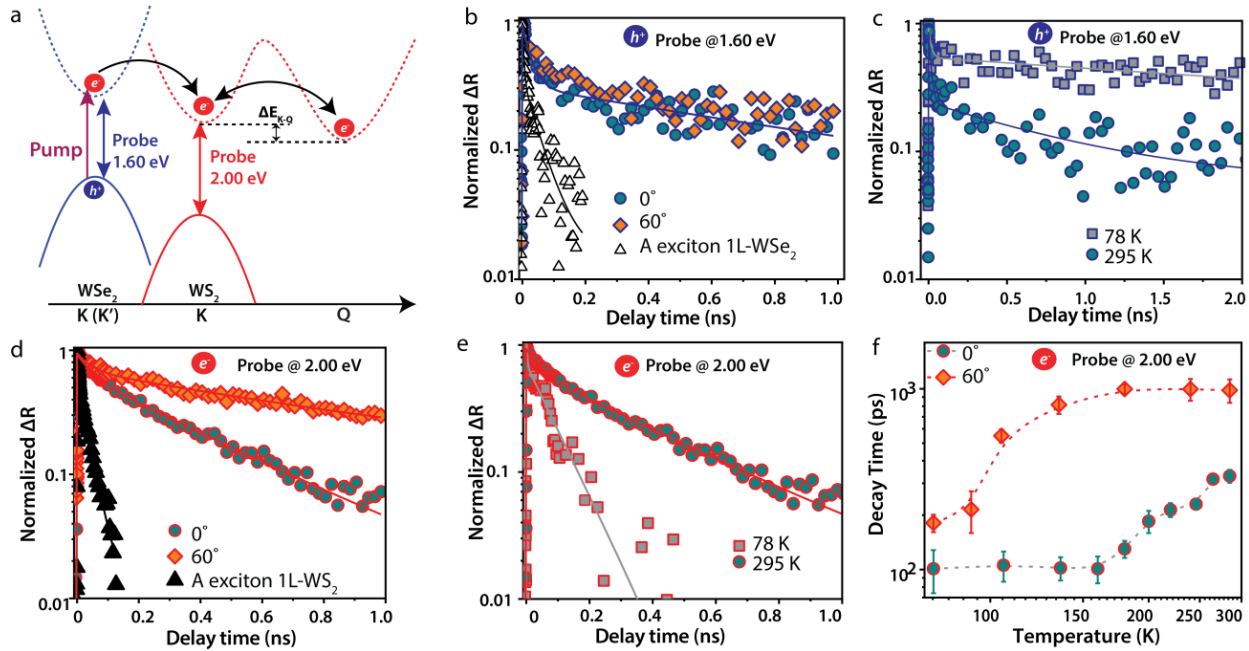

**Fig. 3 | Temperature- and twist-angle-dependent K-K and K-Q interlayer exciton dynamics. a**, Schematic of probing electron and hole dynamics at the K valleys in the WS$_2$-WSe$_2$ heterostructure using transient absorption spectroscopy. The pump-induced change in the probe reflectance ($\Delta R_{pump-on} - \Delta R_{pump-off}$) is collected as function of the pump-probe delay time. The hole dynamics is determined by selectively exciting the WSe$_2$ layer and probing the WSe$_2$ layer using a probe photon energy of 1.60 eV. The hole population corresponds to the sum of K-K and K-Q exciton populations. The electron dynamics is determined by selectively exciting the WSe$_2$ layer and probing the WS$_2$ layer using a probe photon energy of 2.00 eV. The electron population reflects the K-K exciton population only. **b**, Hole dynamics in the heterobilayers with twist angles of 0° and 60° at 295 K. **c**, Temperature-dependent hole dynamics in the 0° heterobilayer. **d**, Electron dynamics in the heterobilayers with twist angles of 0° and 60° at 295 K. Note that the exciton dynamics of the 1L-WS$_2$ is measured using pump and probe energies of 3.10 and 2.00 eV, respectively. **e**, Temperature-dependent electron dynamics for the 0° heterobilayer. **f**, Plot of the fitted decay time of the K-K exciton shown in **e** as function of temperature. The dashed lines are guides to the eye. All measurements were carried out at an exciton density of $4.1 \times 10^{12}$ cm$^{-2}$. All decay curves are fitted using a bi-exponential decay function, convoluted with a Gaussian function. These results support that the K-K and K-Q exciton dynamics are modulated by the twist-angle dependence of the energy gap $\Delta E_{KQ}$.

The twist-angle- and temperature-dependent K-K exciton dynamics can be explained by intraband scattering of electrons between the K and Q valleys, influenced by $\Delta E_{KQ}$. Initially, the electrons transferred from WSe$_2$ reside in the K valley of WS$_2$, but they are quickly scattered to the Q valley by the intervalley scattering (Fig. 3a). The scattering from K to Q through phonon emission does not require thermal activation ($\Delta E_{KQ} > 0$) and is much faster than the electron–hole recombination (~100 ps vs. ~3 ns at 78K). On the other hand, the back scattering of electrons from the Q to the K valley by phonon absorption is a thermal process and the probability of this process is governed by Bose-Einstein statistics, $f(E) \propto 1/(e^{-\frac{\Delta E_{KQ}}{k_B T}} - 1)$, where $k_B$ is the Boltzmann constant and $T$ is the temperature. At higher temperature, more electrons are backscattered to the K valley and therefore $N_{K-K}$ increases, leading to a longer decay time (Fig. 3e). The impact of temperature for the scattering from Q to K can be seen in Fig. 3f, where the K-K exciton decay time increases as the temperature increases. Higher temperature is required for 0° than for 60° to thermally excite phonons to scatter electrons from Q to K (Fig. 3f), directly confirming that $\Delta E_{KQ}$ is larger, as predicted by DFT (88 meV vs. 62 meV, see Table S3 for more details). The larger $\Delta E_{KQ}$ for 0° leads to less efficient Q-K scattering and, hence, shorter K-K exciton decay time than for 60° (Fig. 3d). Therefore, by selectively measuring the dynamics of the K-K and K-Q excitons, these experiments verified prediction (i).

Overall, the PL dynamics is similar to the hole dynamics (Fig. S7), implying contributions from both K-K and K-Q excitons. Because of the 4% lattice mismatch between 1L-WS$_2$ and 1L-WSe$_2$, the K-K excitons have small momentum mismatch and therefore are probably momentum-indirect. The momentum mismatch for K-Q excitons is even larger and should lead to less emission than that from K-K excitons due to the requirement for stronger phonon assistance. At equilibrium, $\frac{N_{K-K}}{N_{K-Q}} = e^{-\frac{\Delta E_{KQ}}{k_B T}}$ and thus, a larger $N_{K-K}$ exists for 60° than 0° at a given temperature. This explains the stronger PL emission intensity and the higher emission energy observed for 60° at 78 K (Fig. 1d), reflecting more emission from K-K excitons.

Next, we investigated the twist-angle dependence of interlayer exciton transport to test our prediction (ii). We have recently demonstrated transient absorption microscopy (TAM) as a new means to image the time-dependent carrier and exciton transport[32,33]. Briefly, a Gaussian pump beam was fixed on the sample and a Gaussian probe beam was scanned relative to the pump beam with a pair of galvanometer scanners, to obtain interlayer-exciton propagation at different time delays (more details can be found in the Supplementary Note 2 and Fig. S12). The interlayer exciton transport in both 60° and 0° heterobilayers was imaged with a linearly polarized pump with a photon energy of 1.60 eV and with a beam size of ~500 nm (Fig. S13), creating A excitons in $WSe_2$. The exciton density at time zero (details in Supplementary Note 3) was between $2.1 \times 10^{12}$ and $6.0 \times 10^{12}$ cm$^{-2}$, corresponding to an mean inter-exciton distance of 4-7 nm, larger than the interlayer exciton radius ($a_0 \sim 2$ nm)[34]. At zero delay time, the excitons have a spatial distribution of $n(x,0) = N_0 exp[-\frac{(x-x_0)^2}{2\sigma_0^2}]$. Within the pulse width of ~200 fs, the electrons were transferred to $WS_2$, leading to the formation of interlayer excitons. We employed a probe energy of 2.00 eV with a beam size of ~400 nm (Fig. S13), to monitor the population of K-K interlayer excitons (no TA signal was observed in the $WS_2$ only region, Fig. S14). How the interlayer excitons move out of the initial volume was measured as function of the pump-probe time, resulting in a population distribution of $n(x,t) = N_t exp[-\frac{(x-x_0)^2}{2\sigma_t^2}]$ (Fig. 4a). The experimentally measured $\sigma_t^2 - \sigma_0^2$ corresponds to the mean squared distance travelled by the interlayer excitons.

We carried out exciton-density-, temperature-, and twist-angle-dependent TAM measurements to directly visualize the localization of interlayer excitons. Normal diffusion would lead to a linear temporal dependence of $\sigma_t^2 - \sigma_0^2$, but a highly nonlinear temporal dependence was observed for the interlayer excitons in both 0° and 60° heterobilayers. The anomalous diffusion with faster transport at higher densities (Fig. 4b and Fig. 4c) can be understood such that the interlayer excitons have permanent electric dipole moments due to the electron–hole separation, which results in repulsive dipole–dipole and exchange interactions, consistent with the previous report in the $MoSe_2$-$WSe_2$ heterostructures[3] and coupled quantum-well heterostructures[23,35,36].

The strong twist-angle dependent of the overall range of anomalous diffusion suggests the role of the moiré potentials. Excitons are more localized for 0° than 60°, with a shorter range of motion for the same exciton density (Fig. 4). While the repulsive interactions between the interlayer excitons have been reported for $MoSe_2$-$WSe_2$ heterostructures[3], a moiré potential was absent from the $MoSe_2$-$WSe_2$ heterostructures with 0° twist angle in that study. Exciton localization by moiré potentials is directly visualized by temperature-dependent measurements (Fig. 4d and Fig. 4e), showing reduced range of exciton motion as temperature decreases. The twist-angle dependence confirms a deeper moiré potential for 0° and validates prediction (ii). The transport of interlayer excitons is significantly faster than that of intralayer excitons in 1L-$WS_2$, 1L-$WSe_2$, 2L-$WS_2$, and 2L-$WSe_2$ for exciton densities of ~$10^{12}$ cm$^{-2}$ (Fig. 4f and Fig. S15), demonstrating the central role of many-body repulsive interactions in the transport of interlayer excitons.

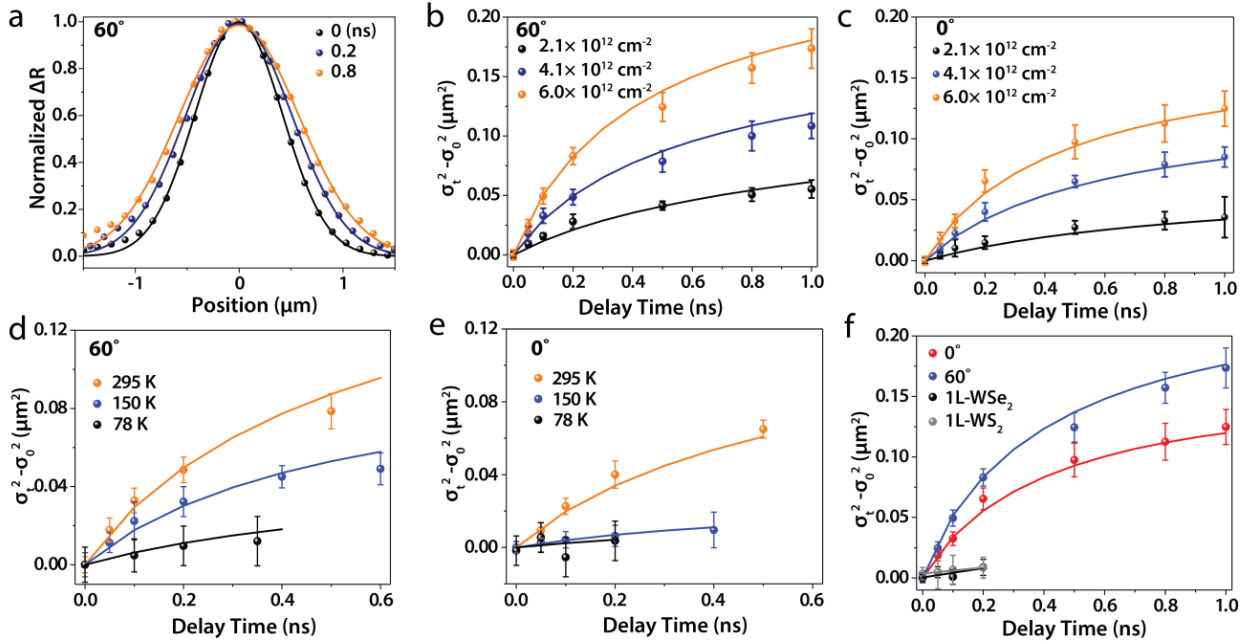

**Fig. 4 | Interlayer exciton transport in moiré potentials. a**, Spatial profiles of the exciton population in WS$_2$-WSe$_2$ (60°) with pump and probe photon energies of 1.6 and 2.0 eV, respectively, at 295 K. The initial exciton density is $6.0 \times 10^{12}$ cm$^{-2}$. **b, c,** $\sigma_t^2 - \sigma_0^2$ as function of the pump-probe delay times at different exciton densities at 295 K for 60° and 0°, respectively. Solid lines are simulations using the Equation 1 described in the main text. **d, e,** Temperature-dependent $\sigma_t^2 - \sigma_0^2$ as function of the pump-probe delay times for 60° and 0° with an initial exciton density of $4.1 \times 10^{12}$ cm$^{-2}$. Solid lines are fits using the Equation 1. **f**, Twist-angle-dependent interlayer exciton transport in the WS$_2$-WSe$_2$ heterobilayers. The interlayer exciton transport in WS$_2$-WSe$_2$ heterobilayers is also compared to the exciton diffusion in 1L-WSe$_2$ and 1L-WS$_2$. The black and grey lines are fits using linear functions. Note that the exciton diffusion in 1L-WS$_2$ and 1L-WSe$_2$ is measured using the pump photon-energy of 3.10 eV and probe photon-energy of 2.00 and 1.60 eV, respectively. These results show that many-body interactions delocalize interlayer excitons in twist-angle-dependent moiré potentials.

To quantitatively describe the motion of interlayer excitons in the moiré potentials, we employed a model that includes the effects of many-body repulsive interactions and the trapping potential [23,35-37]:

$$\frac{\partial n(x,t)}{\partial t} = -\nabla \cdot (J_{Diff} + J_{Rep}) - \frac{n(x,t)}{\tau} \quad (1)$$

where $\tau$ is the exciton lifetime. $J_{Diff}$ describes the exciton diffusion, which can be written as: $J_{Diff} = -D \cdot \nabla n(x,t)$, where $D$ is the diffusion coefficient. $J_{Rep}$ describes the many-body repulsive interaction between the interlayer excitons, which is given by: $J_{Rep} = -n(x,t) \cdot \mu \cdot u_0 \cdot \nabla n(x,t)$, where $\mu$ is the reduced exciton mass and $u_0$ is the exciton interaction energy per unity of density. Because of the small interlayer separation of ~7 Å, the exchange term is much larger than the dipole-dipole interaction and dominates the many-body repulsion force[3,37]. Thus, $u_0$ is given by: $\pi \frac{\hbar^2}{\mu}$ (see the Supplementary Note 4). The diffusion constant $D$ is related to the trapping moiré potential $U$ and can be described as: $D_0 \cdot \exp[-\frac{U}{k_B T + u_0 * n(x,t)}]$, where $D_0$ is the exciton diffusion constant in the absence of the trapping potential [37]. At higher densities, more excitons accumulate in the potential minima and the repulsive exciton-exciton interactions effectively screen the moiré potentials, leading to delocalization of the excitons. Note that Mott density estimated from the Bohr radius[38], $(0.1 - 0.3) \times a_0^{-2}$ =(2.5-7.5)×10$^{12}$ cm$^{-2}$, is similar to the exciton densities at time zero in our measurements. However, recent ultrafast mid-infrared measurements[39] demonstrated strongly bound interlayer excitons with binding energy of ~ 126 meV survived even at density as high as 2 ×10$^{12}$ cm$^{-2}$. Further, as excitons move out the initial pump spot, the exciton density decreases quickly below the Mott density. Therefore, the electron and hole should remain largely correlated and discussion using an exciton picture is valid here.

Thus, the depth of moiré potentials can be determined by modeling the exciton-density- and temperature-dependent exciton transport data. As shown in Fig. 4b and Fig.4d, solid lines represent the simulated $\sigma_t^2 - \sigma_0^2$ as function of the delay time based on Eq. 1 for the 60° heterobilayer, and a moiré potential of $U = 0.11\ eV$ was obtained (other parameters are listed in Table S4). A deeper trapping potential of $U = 0.15$ eV was determined for the 0° heterobilayer by fitting data in Fig. 4c and Fig. 4e. The trapping potentials measured by the TAM experiments represent the averaged potentials experienced by the interlayer K-K excitons, because the pump and probe beam size of ~400 nm is much larger than the moiré periodicity of ~8.5 nm. The measured potential values agree well with the averaged potential heights of 0.08 eV and 0.17 eV for 60° and 0°, respectively, predicted by our DFT calculations (Fig. 2d and Table S2). In the absence of the moiré potentials, the free interlayer excitons are very mobile with a diffusion constant of 9.0 cm$^2$ s$^{-1}$ at room temperature.

The transport of the interlayer exciton is determined by the interplay between the moiré potentials and many-body exciton interactions. Strong many-body interactions can overcome the moiré potentials, leading to a significant delocalization of interlayer excitons at densities >10$^{12}$ cm$^{-2}$. For instance, excitons can travel ~200 nm in 1 ns at a density of 2.0×10$^{12}$ cm$^{-2}$ at room temperature. The delocalization at high densities is consistent with the recent report of broad interlayer PL peak at high density that splits into several narrow lines at low power[15]. PL spectra taken at high exciton densities reflect excitons that sample many locations with different energy levels, resulting in broad emission lines.

The localization and delocalization of the interlayer excitons presented here have important implications for the potential applications of vdW heterostructures; for long-range transport, more delocalized interlayer excitons are preferred and, therefore, deep moiré potentials should be avoided. On the other hand, for applications such as quantum emitters, deep moiré potential should be preferred, to localize excitons. We show that the free interlayer excitons in absence of a moiré potential can be very mobile, with a diffusion constant of 9.0 cm$^2$ s$^{-1}$ at room temperature. Further, interlayer exciton transport significantly deviates from normal diffusion, and therefore to correctly predict exciton diffusion length both the exciton density and the depth of moiré potential have to be taken into account. We also stress that K-Q interlayer excitons are the ground state instead of the commonly assumed K-K excitons and are necessary to be considered when discussing interlayer excitons in the WS$_2$-WSe$_2$ systems.

**Methods**

**Sample fabrication**

The vertically stacked $WS_2$-$WSe_2$ heterostructures were synthesized via a modified two-step chemical vapor deposition (CVD) method[25] schematically shown in Fig. S1. Briefly, a quartz boat with tungsten disulfide powder (99.8%, Alfa Aesar) and a piece of Si wafer with a 285-nm-thick oxide (1 cm × 3 cm) were placed at the center and the downstream of the furnace for the growth of the $WS_2$ monolayers respectively. The system was firstly cleaned by the high pure Ar gas (400 SCCM) for 15 min. The furnace was then heated to 1050 °C and kept for 10 min for the growth of $WS_2$ monolayers. The as-prepared $WS_2$ monolayers were used as the new substrate for the vertically growth of the $WSe_2$ monolayers. The tungsten diselenide powder (99.8%, Alfa Aesar) and the substrate with the $WS_2$ monolayers were placed at the center and downstrem of the quartz tube respectively. A mixture flow of $H_2$/Ar (with 5% $H_2$) gas was used as the carrier gas and the growth temperature was set at 1000°C. After the growth, the furnace was cooled down to the room temperature naturally.

**Scanning transmission electron microscopy**

For scanning transmission electron microscopy (STEM) measurements, CVD grown $WS_2$-$WSe_2$ vertical heterostructures were transferred onto a copper grid using the poly(methyl methacrylate) PMMA-assisted transfer method. The STEM measurements were carried out on JEOL ARM200F microscope operated at 200 kV and equipped with a probe-forming aberration corrector. For HAADF-STEM images, the inner and outer collection angles of the ADF detector are 68 and 280 mrad, respectively. The convergence semiangle is about 28 mard.

**Confocal photoluminescence (PL) microscopy**

PL measurements were performed by employing a home-built confocal PL microscope. A picosecond pulsed diode laser (Pico-Quant, LDH-P-C-450B) with an excitation energy of 2.8 eV and a repetition rate of 40 MHz was used to excite the sample. The laser beam was focused onto the sample using a 50X [(numerical aperture) NA=0.95] objective. The PL emission was collected with the same objective, dispersed with a monochromator (Andor Technology) and detected using a charge coupled device (CCD) (Andor Technology). PL images were acquired using a galvanometer scanner (Thorlabs, GVS012) to scan the excitation beam. Time-resolved PL was measured using a single-photon avalanche diode (PDM series, PicoQuant) and a single-photon counting module (PicoQuant). For temperature-dependent PL measurements, the sample was mounted on a cold finger of a continuous-flow liquid nitrogen cryostat (Janis, ST-500). A 40X (NA = 0.60) objective was used to focus the laser beam onto the sample.

**Transient absorption microscopy (TAM) and spectroscopy**

Transient absorption dynamics and transport measurements were taken using a home-built TAM system, performed in the reflection mode, schematically shown in Fig. S13. Briefly, a Ti: Sapphire oscillator (Coherent Mira 900) pumped by a Verdi diode laser (Verdi V18) was used as the fundamental light source (1.60 eV, 76 MHz, 200 fs). 70% of the pulse energy was fed into the optical parametric oscillator (OPO) (Coherent Mira OPO) to generate the probe beam at 2.00 eV, while the remaining 30% was served as the pump beam. The pump beam was modulated at 1 MHz using an acoustic optical modulator (AOM) (Model R21080-1DM, Gooch&Housego). Both pump and probe beams were spatially filtered. A 60X (NA = 0.95) objective was used to focus both pump and probe beams onto the sample, and the reflection light was then collected by the same objective and detected by an avalanche Si photodiode (Hamamatsu). The change in the probe reflection (Δ*R*) induced by the pump was detected by a lock-in amplifier (HF2LI, Zurich Instrument). For morphological imaging, pump and probe beams were spatially overlapped and a piezo-electric stage (P-527.3Cl, Physik Instrumente) was used to scan the sample with a step size of 200 nm. For exciton transport imaging, a galvanometer scanner (Thorlabs GVS012) was used to scan the probe beam relative to the pump beam in space to obtain the exciton diffusion profiles.

For transient absorption dynamics measurements, the pump and probe beams were spatially overlapped and a mechanical translation stage (Thorlabs, LTS300) was used to delay the probe with respect to the pump. For the measurements of the hole dynamics, we used the output of a high–repetition rate amplifier (Pharos Light Conversion, 400 kHz) to pump two independent optical parametric amplifiers (OPAs), one providing the pump (1.80 eV) and the other supplying the probe (1.60 eV). Both pump and probe beams were spatially filtered. An acousto-optic modulator (Gooch &Housego, R23080-1) was used to modulate the pump beam at 100 kHz. The change in the probe reflection (Δ*R*) induced by the pump was detected by a lock-in amplifier (Stanford Research Instrument, SR830).

For temperature-dependent exciton dynamics and diffusion measurements, the sample was mounted on a cold finger of a continuous-flow liquid nitrogen cryostat (Janis, ST-500). A 40X (NA = 0.60) objective was used to focus both pump and probe beams onto the sample.

**Density functional theory (DFT) calculations**

The moiré potentials were calculated by considering different stacking configurations of the heterobilayers. The configurations were structurally optimized using density-functional theory (DFT) as implemented in the ADF-BAND software[40] (BAND2018, SCM, Theoretical Chemistry, Vrije Universiteit, Amsterdam, The Netherlands, http://www.scm.com). The maximum gradient threshold for geometry optimization was set to $10^{-3}$ hartree $Å^{-1}$. The PBE[41] exchange-correlation functional was used together with the valence triple-zeta polarized (TZP) basis sets composed of Slater-type and numerical orbitals with a small frozen core. Relativistic effects, like spin-orbit coupling (SOC), were taken into account by employing the scalar Zero Order Regular Approximation (ZORA)[42]. The k-space integration was done with the 2D variant of the regular k-space grid using the quadratic interpolation method, resulting in 9 symmetry-inequivalent k-points. The convergence of the k-space grid was checked; a denser grid did not significantly change the transition energies. The van der Waals interactions were accounted for using the D3 correction together with the Becke-Johnson damping function as proposed by Grimme[43].

**Acknowledgments**

L.H. and L.Y. acknowledge the support from U.S. Department of Energy, Office of Basic Energy Sciences through award DE-SC0016356. L.Y. also acknowledges the support from the Purdue University Bilsland Dissertation Fellowship. L.Y. thanks Y. Wan and Z. Guo for their assistance in the instrument development. B.Z. and A.P. acknowledge the National Natural Science Foundation of China (Nos. 51525202, 61574054, 61505051, 61635001). AK, JK, and TB acknowledge ZIH Dresden for computational support. AK thanks the GRK 2247/1 (QM3) for financial support. J.K. acknowledges funding by the German Research Foundation (DFG) under grant no. SE 651/45-1.

**Author contributions**

L.Y. and L.H. designed the experiments. L.Y. carried out the optical measurements. B.Z., A.P., and C.M contributed to sample growth and characterization. L.Y. and L.H. analyzed experimental data. T.B., J.K., and A.K. carried out and analyzed the DFT calculations. L.Y. and L.H. wrote the manuscript with input from all authors.

**Competing interests**

The authors declare that they have no competing interests.

**Data and materials availability**

All data needed to evaluate the conclusions in the paper are present in the paper and/or the supplementary information. Correspondence and requests for materials should be addressed to L.H.

# Supplementary Information for

# Anomalous Interlayer Exciton Diffusion in Twist-Angle-Dependent Moiré Potentials of WS$_2$-WSe$_2$ Heterobilayers

Long Yuan[1], Biyuan Zheng[2], Jens Kunstmann[3], Thomas Brumme[4], Agnieszka Beata Kuc[5,6], Chao Ma[2], Shibin Deng[1], Daria Blach[1] Anlian Pan[2], and Libai Huang[1]

[1]Department of Chemistry, Purdue University, West Lafayette, IN 47907, USA

[2]Key Laboratory for Micro-Nano Physics and Technology of Hunan Province, College of Materials Science and Engineering, Hunan University, Changsha 410082, People's Republic of China

[3]Theoretical Chemistry, Department of Chemistry and Food Chemistry, TU Dresden, 01062 Dresden, Germany

[4]Wilhelm-Ostwald-Institute for Physical and Theoretical Chemistry, Leipzig University, 04103 Leipzig, Germany

[5]Helmholtz-Zentrum Dresden-Rossendorf, Abteilung Ressourcenökologie, Forschungsstelle Leipzig, Permoserstr. 15, 04318, Leipzig, Germany

[6]Department of Physics & Earth Science, Jacobs University Bremen, 28759 Bremen, Germany

**I. Supplementary Notes**

**1. Density functional theory calculations**

The optimized lattice parameters *a* of $WS_2$ and $WSe_2$ monolayers are 3.162 Å and 3.295 Å, respectively. This corresponds to a lattice mismatch of about 4%. For the twist angles of 0° and 60°, this leads to an incommensurate moiré superstructure that could best be approximated by supercells containing thousands of atoms (see Fig. 2). Such large systems are too demanding even for standard DFT. Therefore, we modeled the moiré superstructures by considering the high-symmetry stacking configurations as indicated in Fig. 2 and Fig S7. These configurations were studied by employing primitive cells that contain 6 atoms, with the lattice constant *a* set according to three choices: (i) *a* is equal to the value of the $WS_2$ monolayer (and the $WSe_2$ layer is compressed), (ii) *a* is equal to the value of the $WSe_2$ monolayer (and the $WS_2$ layer is strained), (iii) *a* is an intermediate value that minimized the total strain of the system.

    Fig. 2C shows a schematic representation of a typical band structure of all stacking configurations of the heterobilayer. The band structure of $WS_2$-$WSe_2$ heterobilayer close to the Fermi level is defined by the electron bands (conduction band minimum, CBM) that involve electronic states from the $WS_2$ layer and the hole bands (top of valence band, VBM) that involve states of the $WSe_2$ layer. Therefore, lattice choice (i) was used to determine the band edge energy of the CBM and choice (ii) to determine the band edge energy of the VBM, relative to the vacuum level, respectively. Their difference provides a good estimate of the optical-transition energies (band gaps) of the stacking configurations of the heterobilayer and the effects of the compressive or tensile strain on the electronic structure are minimized. The values of all of these transitions are given in Table S1. The (spin) splitting of the bands at the VBM or CBM Fig. 2C is due to the spin-orbit coupling, giving rise to the total of four interband transitions, out of which,

only the ones from the VBM are indicated by arrows. They are either the momentum-direct transitions in the K valley (K-K) or the momentum-indirect transitions between the VBM in K and the CBM in the Q valley (K-Q). For all stacking configurations, K-Q forms the fundamental band gap.

The values in Figure 2D are given relative to the K-Q transition in the $R_h^X$ stacking, because the moiré potential for the motion of the excitons is defined by the local variations of the transitions within the moiré pattern and not by their absolute values. DFT is known to underestimate the band gap size. However, the band gap differences between different stacking configurations of the same heterobilayer should be well described and, by considering the relative energies, the band gap error cancels out. For that reason, we believe it is not necessary to perform more accurate band structure calculations employing, e.g., hybrid functionals or the GW method.

The interlayer transitions 1 and 2 are either spin-conserving (forming a spin-singlet exciton) or spin-flipping (forming a spin-triplet exciton), depending on the relative lattice orientation (twist angle). In TMDC monolayers, spin-conserving transitions are optically bright and spin-flip transitions are optically dark. However, as recently shown by Yu *et al.*[5], the monolayer selection rules do not apply to heterobilayers, where due to the absence of the in-plane mirror symmetry, both transitions are optically bright. Therefore, we have to consider both transitions on equal footing. The information about the spin-singlet or spin-triplet K-K transitions in Fig. 2D was inferred from the analysis of the spin orientations of the band states in connection with Table 2 of Yu *et al.*[5].

To analyze structural properties of the stacking configurations, we adopted the lattice choice (iii) (see above) and optimized the in-plane lattice parameters (*a*) and the interlayer distances (*d*). These values are given in the following table:

| 0° | | | | 60° | | | |
|---|---|---|---|---|---|---|---|
| System | $\Delta E$ [meV] | $a$ [Å] | $d$ [Å] | System | $\Delta E$ [meV] | $a$ [Å] | $d$ [Å] |
| $R_h^h$ | 109.3 | 3.222 | 3.512 | $H_h^h$ | 0.0 | 3.221 | 2.911 |
| $R_h^X$ | 14.9 | 3.227 | 2.862 | $H_h^X$ | 27.8 | 3.227 | 2.917 |
| $R_h^M$ | 0.0 | 3.226 | 2.858 | $H_h^M$ | 103.5 | 3.226 | 3.463 |

Structural and energetic properties of fully optimized high-symmetry stacking configurations for 0° and 60° twist angles (following the lattice option (iii)): the total energy differences ($\Delta E$), the in-plane lattice parameters (*a*), and the interlayer distances (*d*).

While the in-plane lattice parameters almost do not vary between different stacking configurations, the interlayer distances do.

## 2. Factors that limit the spatial precision of TAM imaging

There are two main sources of noise contributing to the TAM imaging: laser fluctuation noise and electronic noise from the detection system (for example, detector and lock-in amplifier). Noise due to laser intensity fluctuations can be effectively eliminated by using heterodyne lock-in detection with MHz modulation where the intensity of the excitation beam (or additional local oscillator) is modulated by an acoustic-optical modulator[1]. Subsequently, a lock-in amplifier referenced to this modulation frequency can sensitively extract the induced signal. The fluctuation of laser intensity ($1/f$ noise) usually occurs at low frequency (< 10 kHz). When $f$ is in the MHz range, the laser intensity noise becomes near the quantum shot noise limit, which is always present because of the Poissonian distribution of the photon counts at the detector. The pixel dwell time should be significantly longer than the modulation period to allow for reliable demodulation for each pixel. Such a modulation scheme has been successfully applied to transient absorption microscopy to achieve single-molecule sensitivity. In our experiments, we use a modulation frequency of 1 MHz. The TAM instrumentation described here could detect a differential transmission $\Delta T/T$ of $10^{-7}$, three orders of magnitude higher sensitivity than conventional TA spectroscopy.

In the TAM imaging of exciton transport, the measured carrier distribution is convoluted with profiles of pump and probe beams, so that the measured carrier distribution is written as

$$\sigma_{measurement}^2 = \sigma_{exciton}^2 + \sigma_{pump}^2 + \sigma_{probe}^2 \tag{S1}$$

The exciton diffusion is written as

$$L^2 = \sigma_t^2 - \sigma_0^2 \tag{S2}$$

Since $\sigma_{pump}^2$ and $\sigma_{probe}^2$ do not change with the pump-probe delay, the diffusion length is only determined by the change of the exciton density profiles.

Here we performed a sensitivity analysis by differentiating equation S2 and obtain the error of the measured diffusion length written as

$$\Delta L = \sqrt{\frac{\sigma_t^2}{\sigma_t^2 - \sigma_0^2}(\Delta\sigma_t^2)^2 + \frac{\sigma_0^2}{\sigma_t^2 - \sigma_0^2}(\Delta\sigma_0^2)^2} = \sqrt{\Delta\sigma_t^2 + \left(\frac{\sigma_0}{L}\right)^2 (\Delta\sigma_0^2 - \Delta\sigma_t^2)} \tag{S3}$$

From equation (S3), we can clearly see that the error mainly comes from the uncertainty of Gaussian profiles obtained at different time delays, which is determined by the signal-to-noise of the TAM system.

## 3. Determination of exciton density

The absorption of the 1L-WSe$_2$ is determined from the differential reflection spectra. For the micro-reflection spectroscopy, the white light from a stabilized tungsten-halogen light source (Thorlabs) was focused onto a pinhole with a diameter of 10 µm. It was then collimated and focused onto the sample with a 40X (NA = 0.6) objective. The reflected light was collected with the same objective, dispersed with a monochromator (Andor Technology) and detected by a TE cooled charge-coupled device (CCD) (Andor Technology). The differential reflectance is defined as

$$\delta R(\lambda) = \frac{R_{sample} - R_{substrate}}{R_{substrate}} \tag{S4}$$

where $R_{sample}$ is the reflectance intensity of the sample with substrate and $R_{substrate}$ is the reflectance intensity of the bare substrate. For temperature-dependent measurements, the sample was mounted on a cold finger of a continuous-flow liquid nitrogen cryostat (Janis, ST-500).

For the ultrathin film on a transparent substrate, the differential reflection is directly related to the absorption by the following equation[2-4]:

$$\delta R(\lambda) = \frac{4}{n_{sub}^2 - 1} A(\lambda) \tag{S5}$$

where $n_{sub}$ is the refractive index of the quartz substrate. We assume it to be wavelength independent for the spectral range investigated in this study. The calculated absorption of 1L-WSe$_2$ at different temperature is shown in Fig. S16.

In the TAM measurements, excitons are generated by the absorption of pump lasers. The peak fluence of pump pulse ($P_f$) could be calculated as:

$$P_f = \frac{P}{A*f} \tag{S6}$$

where $P$ is the average laser power, $A$ is the effective area of the pump beam, $f$ is the repetition rate of the laser. Then, the injected exciton density ($n$) is obtained as:

$$n = \frac{A(\lambda)*P_f}{\hbar\nu} \tag{S7}$$

where $A(\lambda)$ and $\hbar\nu$ are the absorption and photon energy of pump pulse respectively.

## 4. Exchange and dipole-dipole interactions

For small transferred momenta in exciton-exciton scattering, exciton interaction energy per unity of density ($u_0$) is given by

$$u_0 = \pi \frac{\hbar^2}{\mu \chi(d)} \tag{S8}$$

where $\mu$ is the reduced exciton mass and $d$ is the effective separation between the electron and hole layer. The dimensionless function $\chi(d)$ has two-well defined limits, $d \ll a_{2d}^B$ and $d \geq a_{2d}^B$ (2D exciton Bohr radius). The first limit corresponds to the exciton-exciton exchange interaction. In this situation, $\chi(d \ll a_{2d}^B)$ is very close to unity[6], $\chi(d) = 1.036$, according to first-principles calculations[7]. The second limit is a well-defined dipole-dipole interaction. In this case, $\chi(d > a_{2d}^B) = a_{2d}^B/(2d)$ and eq. (S8) reduces to $u_0 = 4\pi(e^2/\varepsilon)d$, where $\varepsilon$ is the dielectric constant.

## II. Supplementary Figures

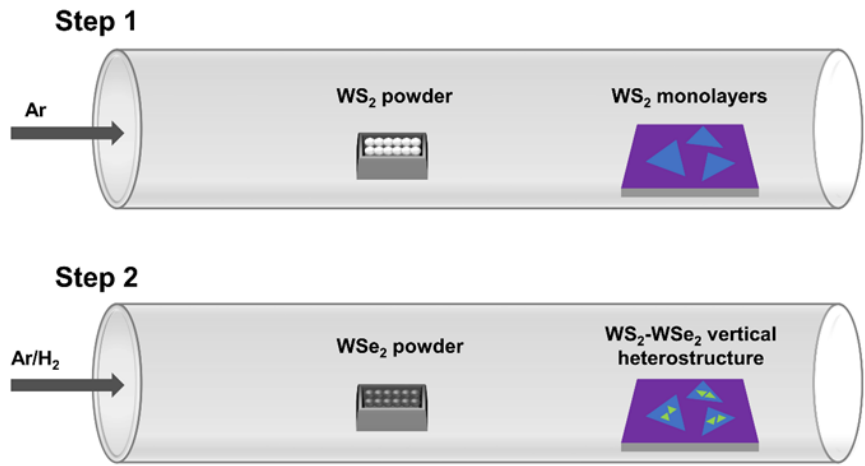

**Fig. S1.** Schematic of the $WS_2$-$WSe_2$ vertical heterostructures prepared using the modified two-step CVD method.

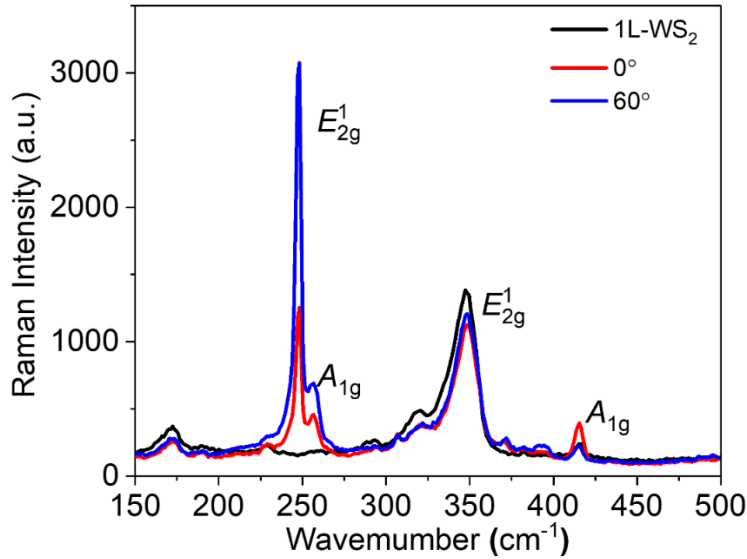

**Fig. S2**. Raman spectrum of the 1L-WS$_2$, WS$_2$-WSe$_2$ (0°), and WS$_2$-WSe$_2$ (60°). In the case of the 1L-WS$_2$, the $E_{2g}^1$ and $A_{1g}$ modes are determined to be ~ 350 cm$^{-1}$ and ~ 420 cm$^{-1}$ respectively. For the 1L-WSe$_2$, $E_{2g}^1$ and $A_{1g}$ modes are observed at ~ 250 cm$^{-1}$. The Raman peak position and intensity ratio of the 1L-WS$_2$ and 1L-WSe$_2$ are consistent with the previous report[8]. The Raman spectra were collected by a confocal Raman system (HORBIA Jobin Yvon, XploRA). A CW laser with an excitation energy of 2.33 eV was used to excite the sample, which was focused onto the sample with a 100X (NA= 0.90) objective. A grating of 1800 lines per mm was used in the measurements which provides a spectral resolution of 1.9 cm$^{-1}$. To minimize any thermal damage in the air, all measurements were taken with excitation power of 1 mW and an integration time of 1 s. The spectrometer was calibrated with a Si substrate at 520 cm$^{-1}$ before the measurements.

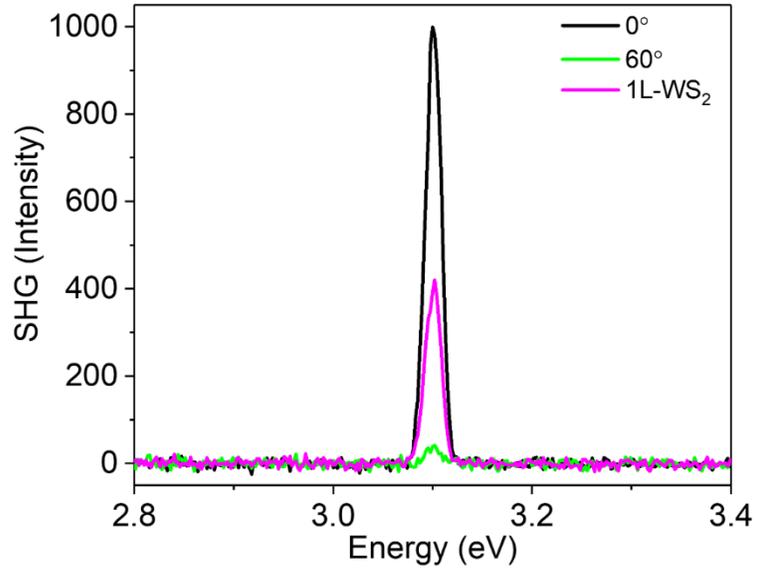

**Fig. S3.** SHG measurements of the 1L-WS$_2$, WS$_2$-WSe$_2$ (0°), and WS$_2$-WSe$_2$ (60°). The SHG measurements were taken using a home-built inverted microscopy system. A linear polarized femtosecond laser (1.55 eV) from an optical parametric amplifier (OPA, TOPAS-Twins, Light Conversion Ltd) pumped by the output of a high repetition rate amplifier (PHAROS Light Conversion Ltd., 400 KHz, 200 fs) was focused onto the sample with a 50X (NA = 0.95) objective. The reflected SHG light (3.10 eV) was collected using the same objective, dispersed with a monochromator (Andor Technology) and detected by a TE cooled charge-coupled device (CCD) (Andor Technology).

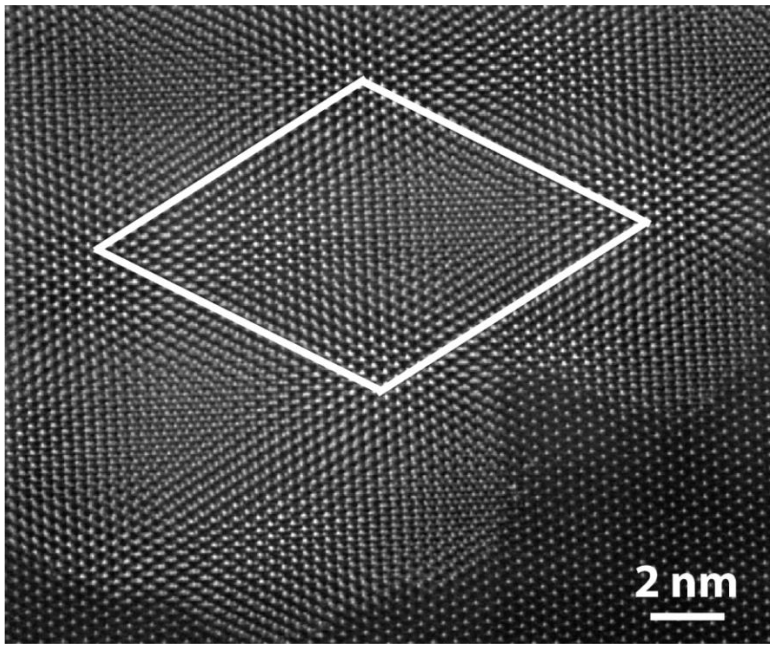

**Fig. S4.** STEM image of the WS$_2$-WSe$_2$ (0°). The moiré superlattice is marked by white solid lines showing a periodicity of ~8.5 nm.

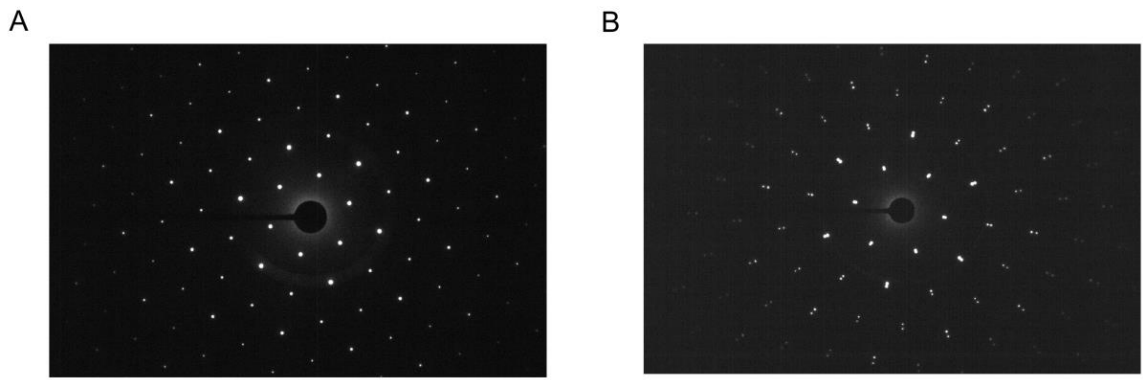

**Fig. S5.** Selected-area electron diffraction (SAED) of the 1L-WS$_2$ and WS$_2$-WSe$_2$ (60°) which confirms the lattice mismatch of the WS$_2$ and WSe$_2$ monolayers.

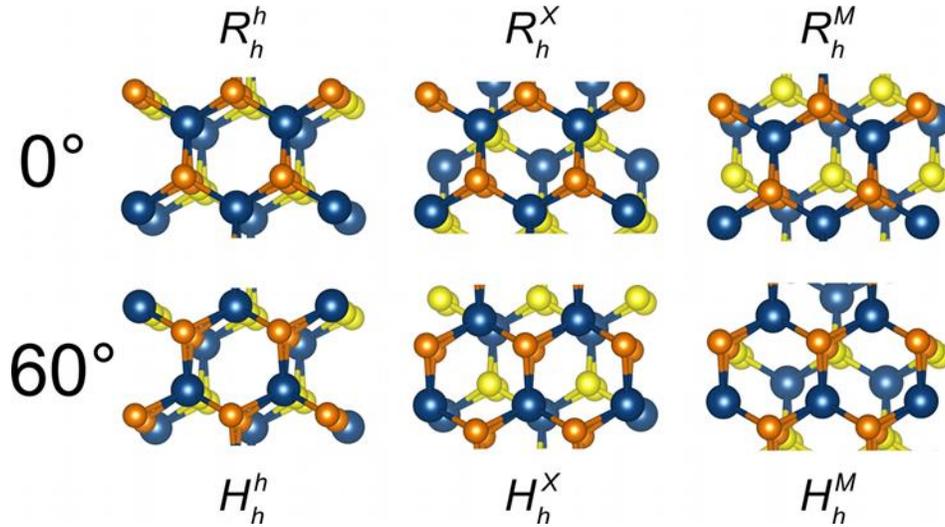

**Figure S6.** High-symmetry stacking configurations for the twist angles 0° (top) and 60° (bottom), as considered in the DFT analysis.

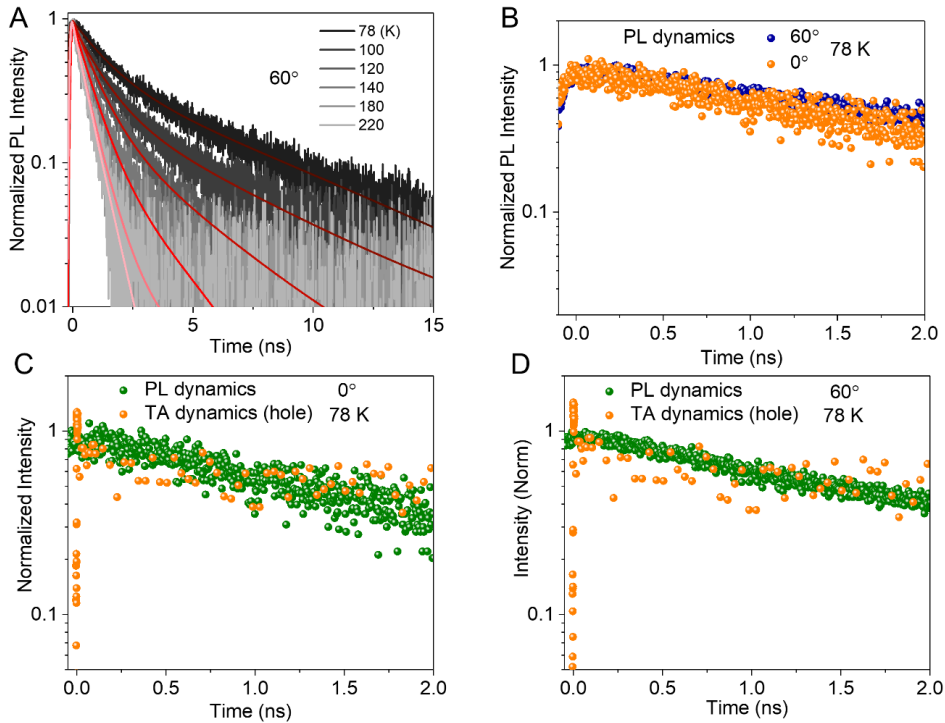

**Fig. S7.** (A) Temperature-dependent interlayer exciton PL dynamics of the $WS_2$-$WSe_2$ (60°) showing that the PL lifetime of interlayer exciton increases as temperature decreases. (B) PL dynamics of the $WS_2$-$WSe_2$ (0°) and $WS_2$-$WSe_2$ (60°) at 78 K displaying twist-angle-independent behavior which is consistent with TA hole dynamics. PL dynamics versus TA dynamics of the holes in (C) $WS_2$-$WSe_2$ (0°) and (D) $WS_2$-$WSe_2$ (60°). We note that the initial fast components in the TA dynamics of the holes arise from the A exciton decay of the 1L-$WSe_2$ because pump excites the $WSe_2$ and the slow components reflect the interlayer exciton recombination.

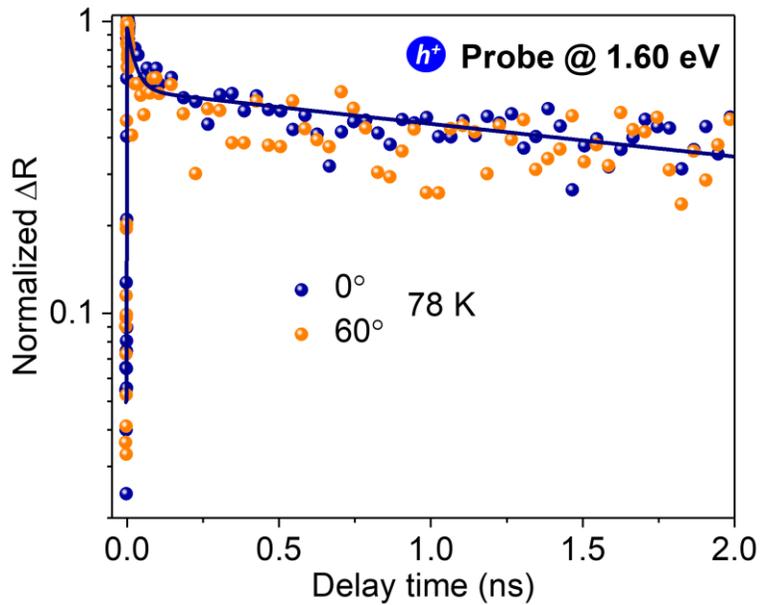

**Fig. S8.** Hole dynamics in the $WS_2$-$WSe_2$ heterostructures with twist angles of 0° and 60° at 78 K. Pump and probe photon energies are 1.80 and 1.60 eV respectively. The solid line is a fit using a bi-exponential decay function convoluted with a Gaussian function.

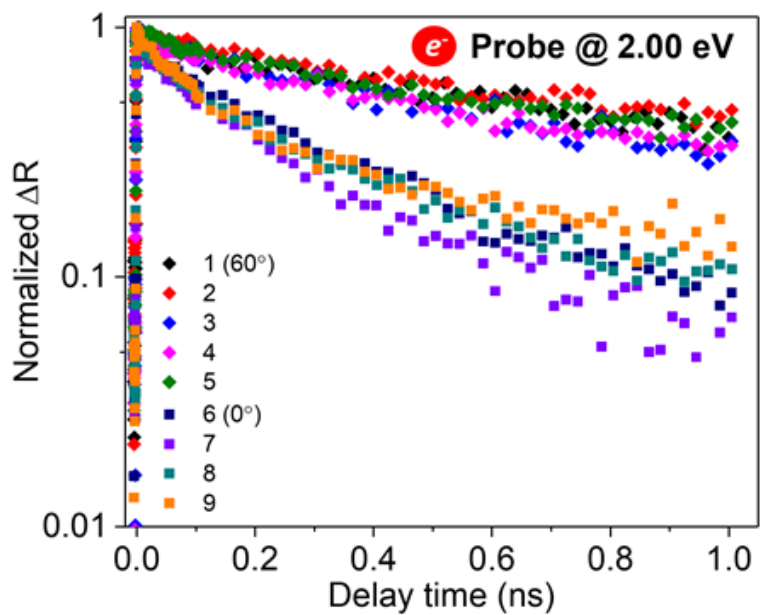

**Fig. S9.** Electron dynamics of the $WS_2$-$WSe_2$ heterostructures with twist angles of 60° and 0° in different samples at 295 K. We observed similar twist-angle dependent electron dynamics in different samples.

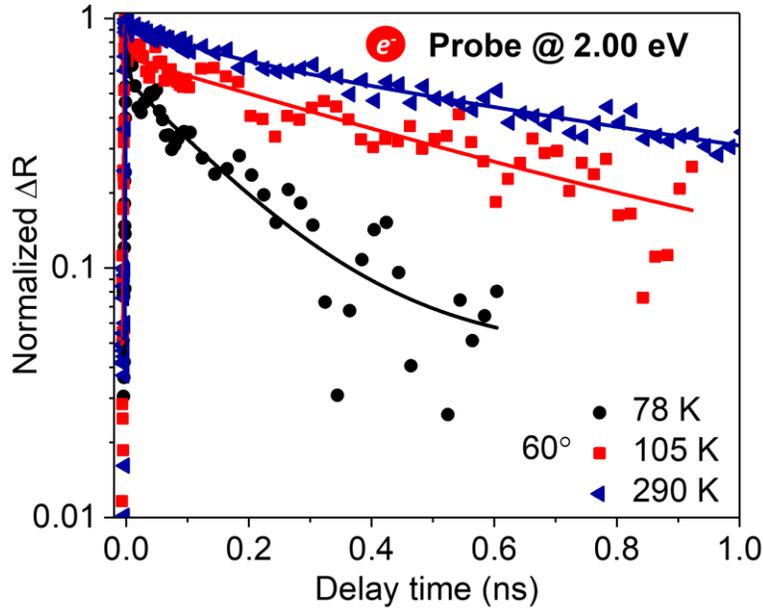

**Fig. S10.** Temperature-dependent electron dynamics of the WS$_2$-WSe$_2$ (60°) with pump and probe photon energies of 1.60 and 2.00 eV respectively. Solid lines are fits using a bi-exponential decay function convoluted with a Gaussian function.

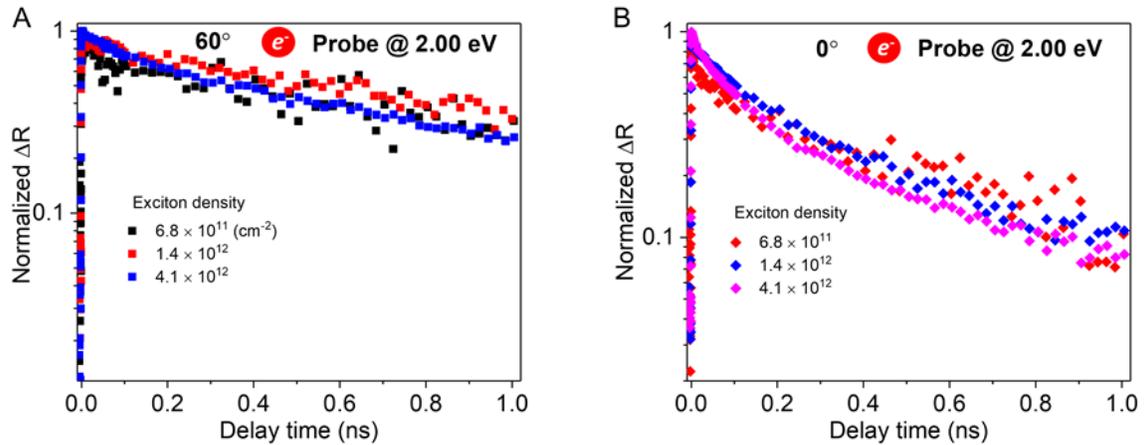

**Fig. S11.** Exciton-density dependent electron dynamics of the WS$_2$-WSe$_2$ heterostructures with twist angles of 60° (A) and 0° (B) at 295 K.

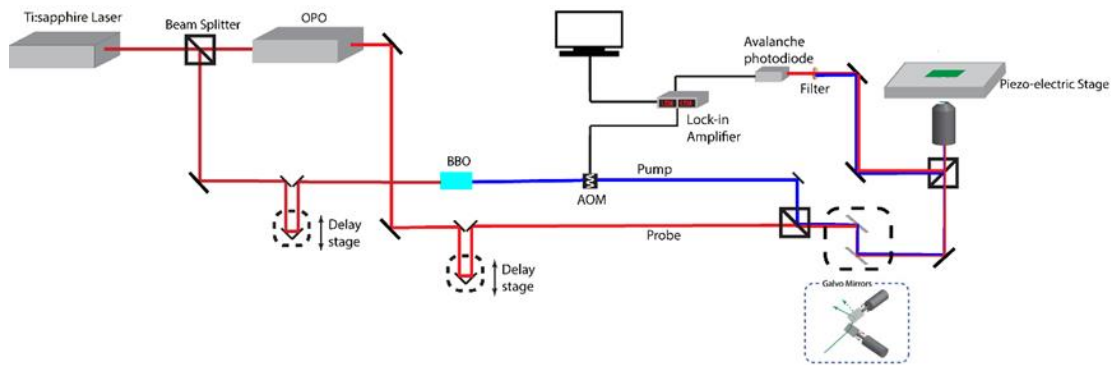

**Fig. S12.** Schematic of a home-built TAM setup (reflection mode). OPO: optical parametric oscillator; AOM: acoustic optical modulator; BBO: beta barium borate.

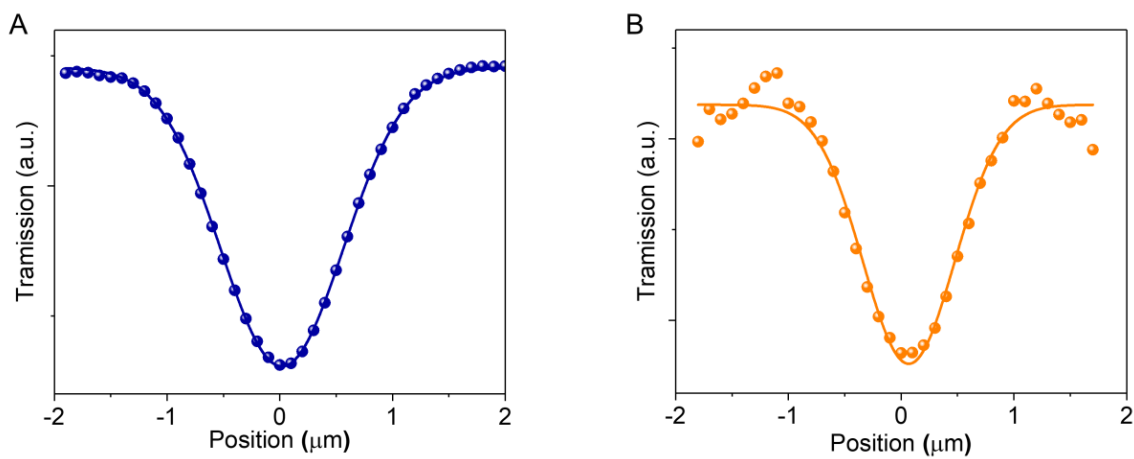

**Fig. S13.** We experimentally measured the laser beam size by scanning the beam across a thin gold nanowire with a diameter ~ 50 nm. By fitting the spatial profiles with Gaussian functions, we determined the width ($\sigma$) of laser beam with photon energies of 1.6 (A) and 2.0 eV (B) to be ~0.5 and 0.4 µm respectively.

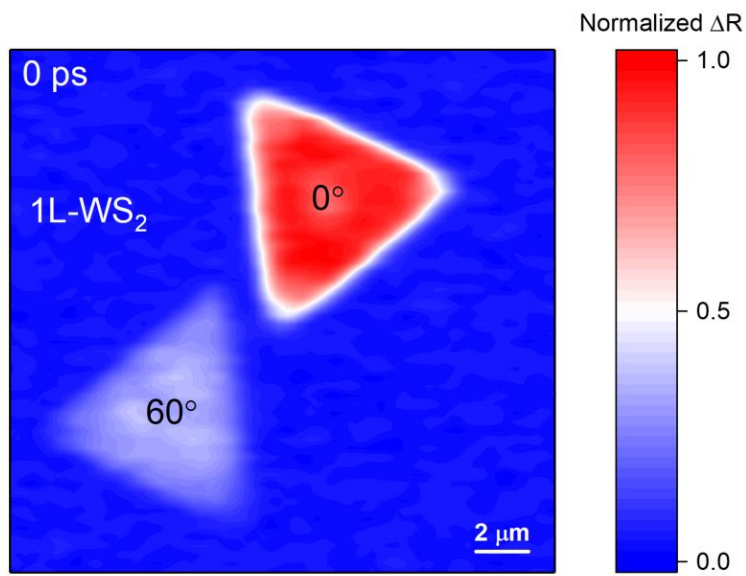

**Fig. S14.** TAM image of the WS$_2$-WSe$_2$ (0°) and WS$_2$-WSe$_2$ (60°) at 0 ps when both pump and probe beams are spatially overlapped. The pump and probe photon energies are 1.60 and 2.00 eV respectively. It displays that the WS$_2$ only area has no detectable TA signal due to the pump photon energy (1.60 eV) is well below the bandgap of the WS$_2$ (2.00 eV).

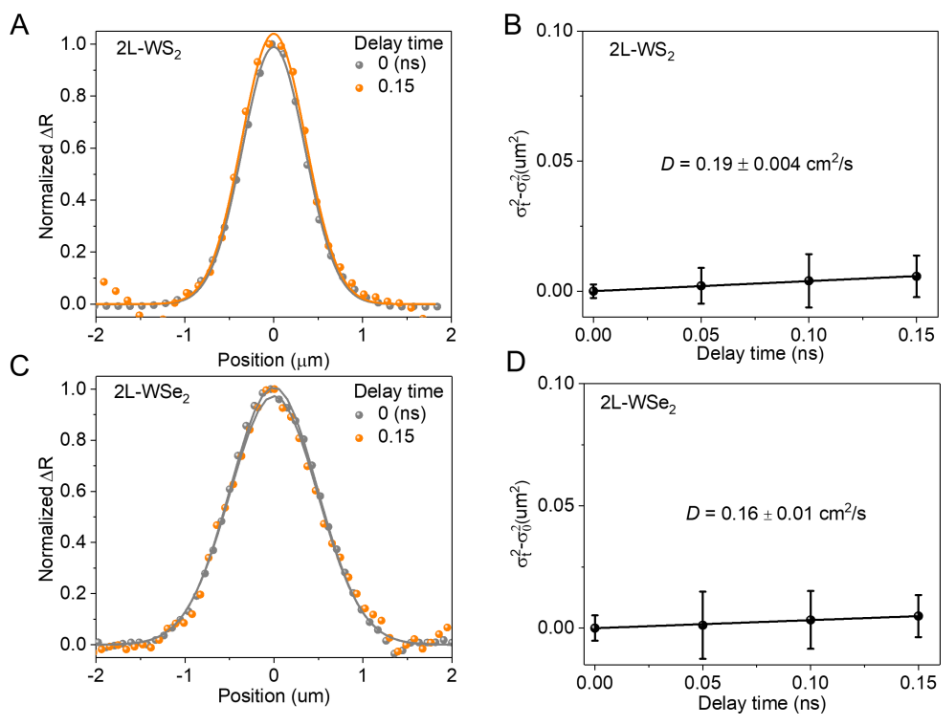

**Fig. S15.** Exciton diffusion in the 2L-WS$_2$ and 2L-WSe$_2$ at 295 K. The pump photon energy is 3.10 eV. The probe photon energies are 2.00 eV and 1.60 eV for 2L-WS$_2$ and 2L-WSe$_2$ respectively. The exciton density is ~ $6.0 \times 10^{12}$ cm$^{-2}$.

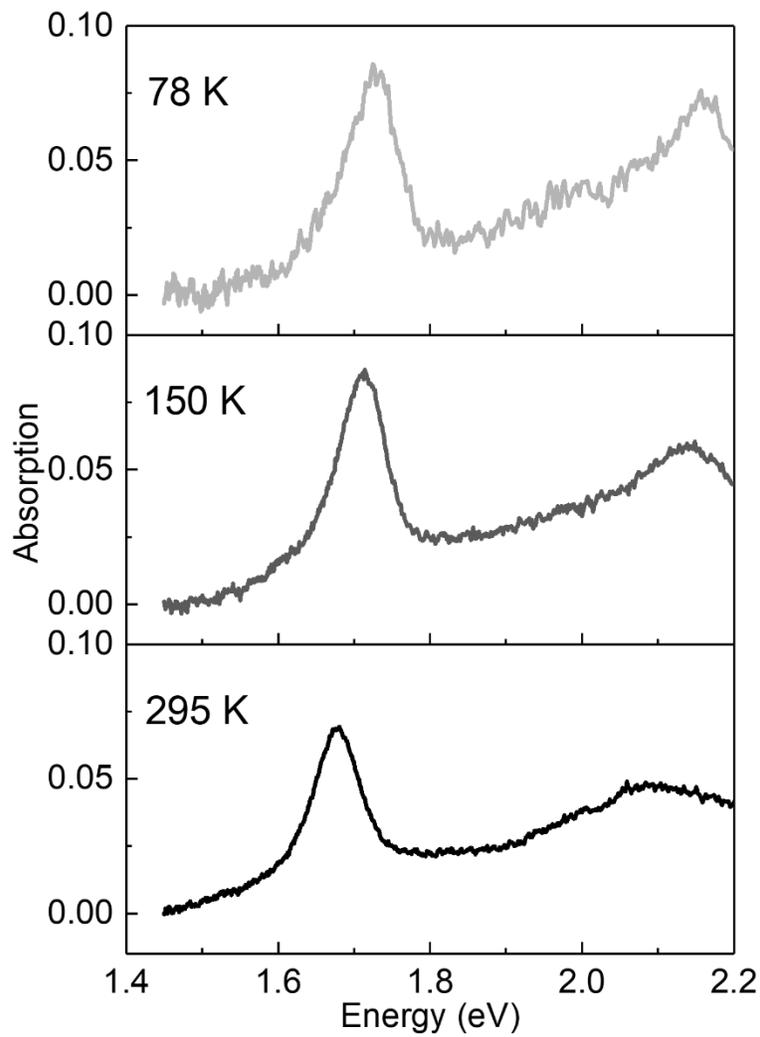

**Fig. S16.** Temperature-dependent absorption spectrum of the 1L-WSe$_2$.

## III. Supplementary tables

| k-space | Transition | 0° | | | 60° | | |
|---|---|---|---|---|---|---|---|
| | | $R_h^h$ | $R_h^X$ | $R_h^M$ | $H_h^h$ | $H_h^X$ | $H_h^M$ |
| K—K | 1 | 0.978 | 0.949 | 1.151 | 1.056 | 1.044 | 0.987 |
| | 2 | 1.015 | 0.995 | 1.186 | 1.096 | 1.078 | 1.024 |
| K—Q | 3 | 0.919 | 0.869 | 1.026 | 0.948 | 1.008 | 0.946 |
| | 4 | 1.120 | 1.075 | 1.204 | 1.024 | 1.087 | 1.073 |

**Table S1.** Calculated energies of the K-K and K-Q optical transitions for high-symmetry stacking configurations (in eV). The four values for K-K and K-Q correspond to the transitions indicated in Fig. 2C.

| Transition | 0° | 60° |
|---|---|---|
| 1 (K—K) | 0.202 | 0.072 |
| 2 (K—K) | 0.191 | 0.069 |
| 3 (K—Q) | 0.157 | 0.062 |
| 4 (K—Q) | 0.129 | 0.063 |
| Average | **0.170** | **0.067** |

**Table S2.** Heights of the moiré potentials (difference between the maximum and the minimum values) for the K-K and K-Q transitions in Table S1 and Fig. 2D. Indices refer to the transitions as indicated in Fig. 2C. The results clearly indicate that the potential height for 0° is much larger (steep potential) than for 60° (shallow potential).

| 0° | 1(K-K) — 3(K-Q) | 60° | 1(K-K) — 3(K-Q) |
|---|---|---|---|
| $R_h^h$ | 0.059 | $H_h^h$ | 0.108 |
| $R_h^X$ | 0.080 | $H_h^X$ | 0.036 |
| $R_h^M$ | 0.125 | $H_h^M$ | 0.041 |
| Average | **0.088** | Average | **0.062** |

**Table S3.** Calculated energy differences (in eV) between the Q and the K valley at the conduction band edge for different stacking configurations, as indicated in Fig. 2C. This also corresponds to the shaded area in Fig. 2D. After averaging over the moiré potential, it is found that this energy difference is 26 meV larger for 0° than for 60°.

|  | $T$ (K) | $D_0$ (cm$^2$/s) | $\tau$ (ns$^{-1}$) | $U$ (eV) |
|---|---|---|---|---|
| WS$_2$-WSe$_2$ (0°) | 295 | 9.0 | 0.9 | 0.15 |
|  | 150 | 4.6 | 0.54 | 0.15 |
|  | 78 | 2.3 | 0.27 | 0.15 |
| WS$_2$-WSe$_2$ (60°) | 295 | 9.0 | 1.2 | 0.11 |
|  | 150 | 4.6 | 0.46 | 0.11 |
|  | 78 | 2.3 | 0.32 | 0.11 |

**Table S4.** Fitted parameters for the temperature-dependent interlayer exciton transport shown in Fig. 4D and Fig. S16. The exciton density is $4.1 \times 10^{12}$ cm$^{-2}$.